\documentclass[12pt]{article}
\usepackage{amsmath}
\usepackage{amssymb}
\usepackage{amscd}
\usepackage[dvips]{graphicx}
 \linethickness{0.4pt}
\begin{document}

\begin{center}
\section*
{High spin particles with spin-mass coupling II}
\end{center}
\begin{center}
{M. DASZKIEWICZ}
\end{center}
\begin{center}
{Institute of Theoretical Physics, Wroc{\l}aw University, pl. Maxa
Borna 9\\
50-206 Wroc{\l}aw, Poland \\
marcin@ift.uni.wroc.pl}
\end{center}
\begin{center}
{Z. HASIEWICZ, T. NIKICIUK and C. J. WALCZYK}
\end{center}
\begin{center}
{Institute of Theoretical Physics, University of Bia{\l}ystok, ul.
Lipowa 41\\
15-424 Bia{\l}ystok, Poland} \\
zhas@uwb.edu.pl, niki@alpha.uwb.edu.pl, c.walczyk@alpha.uwb.edu.pl
\end{center}

\begin{abstract}
    The classical and quantum model of high spin particles within the manifestly
    covariant framework. The internal (spin) degrees of freedom are described by two
    ${\cal C}(3,1)$ Clifford algebra spinors. The covariant quantization leads to PCT
    invariant  spectrum of particles with
    spin dependent masses. The quantum model contains elementary particles and the
    cluster states generating infinite degeneracy of the mass spectrum.

\end{abstract}

\section{Introduction}
There are many possible descriptions of the spinning particles with
arbitrary spin spectrum. It seems, that the most promising ones are
so called spinorial models \cite{BZ}-\cite{andrzej} with the
internal degrees of freedom (spin) realized by classical,
(anti-)commuting spinors, i.e the elements of irreducible or
reducible representation modules of corresponding Clifford algebras,
most commonly ${\cal C}(3,1)$. Despite of the fact that their
construction relies on direct generalization of superparticle models
they contain the particles of arbitrarily high spins. However, the
masses of these particles appear to be the same once the value of
the mass parameter of the model is fixed. There are also
constructions \cite{KLS} which lead to spin dependent masses of the
particles but, in contrast to the models mentioned above, with spin
variable introduced as a free parameter. \\
From this point of view the classical model defined in \cite{HDS}
and analyzed in details in \cite{DHW} seems to be particulary
interesting. It's construction relies on minimal coupling of the
particle trajectory with  vector current build up of single
${\cal C}(3,1)$ real (Majorana) spinor.
Due to this
coupling the resulting first quantized model describes the infinite
families of spinning particles with spin-dependent masses. As the
analysis at the classical level strongly indicates this is this very
coupling, which corresponds to, and is responsible for spinning
particles "zitterbewegung" phenomenon \cite{schroedinger},\cite{BB},\cite{rivas},
which is most clearly and efficiently described in Clifford algebra language. \\
Besides of the "dynamical" spin-mass coupling, the spinning particle
model of \cite{HDS} and \cite{DHW} has an additional advantage: the
non degenerate mass levels.  Unfortunately, it has also important
drawback. The first quantization results in the Hilbert space of
states, which contains orphaned\footnote{The corresponding
antiparticles are missing.} particles and antiparticles, i.e. it
breaks CPT symmetry \cite{Jost}. Consequently, it does not allow one
to construct the corresponding local quantum
field theory by direct application of the second quantization procedures \cite{Jost}.\\
\\
The main task of this paper is to present the extended spinorial
model which cures the above weakness of the prototype proposed in
\cite{DHW}. The appropriate modification consists in adding an
additional spinorial degree of freedom with opposite spin-mass
coupling constant. The model obtained in this straightforward way
contains however the non-linear interaction of spinors. This
interaction is eliminated by the substraction from the Lagrange
function  the quartic term in spinorial variables (see (\ref{int})).
As a result, one gets after first quantization, the infinitly
degenerate but CPT invariant
spectrum of particles. \\
As in the case of \cite{DHW} the paper contains also covariant
formulation of the model based on the complex polarization of the
second class constraints and application of Gupta-Bleuler
quantization procedure \cite{GB}. The polarization leads to the
description of the internal degrees of freedom in terms of Weyl variables i.e.
in the language of even subalgebra ${\cal C}(3,1)_0 \approx {\cal C}(3,0)$ spinors.
This approach results in a natural
generalization of Dirac type equations \cite{Lop} for
the particles with "dynamical" spin-dependent mass.\\
\\
Since the above model has CPT invariant mass spectrum, it can serve
as a starting point for the construction of the local quantum field
theory \cite{Jost} for particles lying on Regge trajectory. It would
allow one, in particular, to investigate the different kinds of
field theoretical interactions of the spinning particles with
arbitrary spin and with spin dependent mass in four dimensional
space-time. \\
As far as we know, the quantum field theoretical description of the
particles lying on Regge trajectories was proposed many years ago in
the series of papers
\cite{field1},\cite{field2},\cite{field3},\cite{field4}. Apart of
that, such a description was also elaborated in the framework of
dual model \cite{Rebbi} and string theory
\cite{hikko},\cite{witten}, but only for the critical dimension of
space-time. In the case of dimension four, the direct constructions
of this kind encounter serious
difficulties and problems related to the presence of Liouville modes \cite{Polyakov}. \\
\\
The paper is organized as follows. \\
In the first chapter the structure of the classical model is
presented in Lagrangian and Hamiltonian form. The complex
parametrization of the spinor degrees of freedom in terms of Weyl
spinors is introduced in order to make the transparent link with the
commonly
used formalism. \\
The next chapter is devoted to the description of particle content
of the corresponding quantum theory. The spectrum is divided into
two series: the elementary trajectories and the sector describing
clusters. \\
Finally, the results are summarized and some open questions and
problems are raised.


\section{The classical model}
The classical model considered in this paper is a natural
generalization and extension of the one presented in \cite{DHW}. The
particle content of the model of \cite{DHW} appeared to be not CPT
invariant:  the particles and antiparticles were orphaned. \\
From the form of the mass spectrum obtained in \cite{DHW}: $m_j =
\sqrt{h^2j^2+m_0^2} - hj$ for the particles of spin $j$ and $m_j =
\sqrt{h^2j^2+m_0^2} + hj$ for antiparticles, one may easily conclude
that in order to get rid of this undesired orphancy, it is enough to
add  an additional spinorial degree of freedom with opposite
spin-mass coupling. The classical model obtained in such a way was
already proposed in the Conclusions of \cite{DHW}. It is defined by
the following Lagrange function:
\begin{equation}
\label{action}
    {\cal L}_{0}=\frac{1}{2}e^{-1}\dot{x}^2-\frac{1}{2}e
    m_{0}^{2} + \frac{h}{2} \dot{x}\cdot j_{1} - \frac{h}{2} \dot{x}\cdot j_{2} +
    \bar{\eta}_{1}{\dot \eta}_{1}
    + \bar{\eta}_{2}{\dot \eta}_{2}\;,
\end{equation}
which is a straightforward extension of the one presented in
\cite{DHW}. In addition to the standard terms describing the scalar
particle of mass $m_0$ it contains "minimal" couplings of the
velocity with two spinor currents
\begin{equation}
\label{currents_two}
j^{\mu}_{I} =
\bar{\eta}_{I}\gamma^{\mu}{\eta}_{I}\;,\;I=1,2\;.
\end{equation}
The currents are
build out of two independent Majorana spinors
${\eta}_{I}$. For the Majorana spinors to exist one should
assume the spacetime metric in the form $g = {\rm
diag}(-1,+1,+1,+1)$ i.e. the Clifford algebra generators present in
(\ref{currents_two}) should satisfy:
$$
\gamma^\mu \gamma^\nu + \gamma^\nu\gamma^\mu = 2 g^{\mu\nu}\,.
$$
The spinorial conjugation is defined in the standard way:
$\bar{\eta}_{I} = {\eta}_{I}^{{\rm T}}\gamma_0$ and  it defines the
${\mathbf{Spin}} (3,1)$ invariant scalar product: $\bar{\eta}\eta '
= {\eta}_{I}^{{\rm T}}\gamma_0\eta ' \,$. The explicite form of this
product is not in fact needed. The only property which is used
further on is that it generates $\beta_-$ antiautomorphism of
${\cal C}(3,1)$ Clifford algebra: $\beta_- (\gamma^\mu) = - \gamma^\mu\,$.\\
The classical model defined by (\ref{action}) appears to generate
non-linear classical equations of motion.\\
    Besides of the standard algebraic equation or $1$-bein variable
    $e$ and momentum conservation law:
$$
\frac{d}{d\tau}\, p^\mu = 0\;,\;\;\; {\rm where}\;\;\; p^\mu =
    \frac{1}{e}\dot{x}^{\mu} + \frac{h}{2} j^\mu_1 - \frac{h}{2}
    j^\mu_2\;,
$$
it contains non-linear equations of motion for spinor variables:
    \begin{equation}
    \label{spinoreqns}
\frac{d}{d\tau}\, \eta_{1/2} = \frac{eh}{2} \left( \mp p^\mu -
\frac{h}{2} j^\mu_{2/1}\right)\gamma_{\mu}\,\eta_{1/2}
    \end{equation}
The interactions of this type are absent in single spinor model of
\cite{DHW} as the spin vector currents built out of single ${\cal
C}(3,1)$ Clifford algebra spinors are light-like. This fact can be
easily proved by the use of Clifford algebra structural relations
without any use of explicit matrix representation.
    This non-trivial "interaction" of spinors follows from the fact
    that both spinors are coupled to single particle trajectory.\\
    For this reason the model defined by (\ref{action}) is difficult
    to treat technically at both: classical and quantum levels. \\
    The form of the equations of motion (\ref{spinoreqns}) strongly
    suggests that the interaction of spinors is of potential
    character. It is then sensible to ask for the corresponding
    linearized, free model. It appears that it can be obtained by
    adding to (\ref{action}) the term which describes the
    cross interaction of spinor currents:
\begin{equation}
\label{int}
    {\cal L}_{\rm{int}} = -\frac{eh}{4}^2j_{1}\cdot j_{2}\;.
\end{equation}
It is not difficult to check (using the property of spinor currents
$j_{I}^2=0$ being null and Schwartz inequality) that (\ref{int}) is
non-negative. This means that the interaction present in
(\ref{action}) and in (\ref{spinoreqns}) is governed by non-negative
potential of fourth order.
    As it appears the term (\ref{int}) cancels the non-linear
interactions at the Hamiltonian level and one is left with
relatively simple system of constraints of mixed type. On the other
hand its presence leads to essential modification of the proper time
parameter. As it can be calculated from the equations of motion for
$1$-bein $e$, the corresponding density gets rescaled:
\begin{equation}
\label{einbein}
    ds = \sqrt{-\dot{x}^2}d\tau \;\;\rightarrow\;\;
    ds' = \sqrt{1 + \frac{h^2j_{1}\cdot j_{2}}{2m_0^2}}
    \,\sqrt{-\dot{x}^2}d\tau\;.
\end{equation}
The questions related to the above change of the internal geometry
of the particle trajectory are by all means interesting, but they
will be not pursued here. The considerations of this paper will be
focused on the hamiltonian formulation and the corresponding first
quantized theory.\\
\\
    At the canonical level the model defined by the sum of (\ref{action}) and
(\ref{int}) describes the constrained system  with constraints of
mixed type. Their structure is analogous  to that
of \cite{DHW}. \\
There are two families of spinorial constraints:
\begin{equation}
\label{secondclasscon}
    G_{I}^{\alpha} = \pi_{I}^{\alpha} + \eta^{\alpha}_I\;\;\;;\;\;\alpha =
    1,...,4\;;\;I=1,2\;,
\end{equation}
relating the canonical momenta $\pi_{I}^{\alpha}$ of spinors with
spinors $\eta^{\alpha}_I$ themselves\footnote{One obviously assumes
the canonical Poisson brackets $\{ \pi_{I}^{\alpha},\eta^{\beta}_J
\}\ = \delta_{IJ}C^{{\alpha}{\beta}}$. The matrix
$C^{{\alpha}{\beta}}$ is the inverse of the one defined in the
following way: $\bar{\eta}_{I}{\eta}_{I}' = {\eta}^{{\rm
    T}}_{I}\gamma_{0}{\eta}_{I}' = {\eta}^{\alpha}_{I} C_{\alpha \beta}{\eta
    '}^{\beta}_{I}$.}. \\
These constraints reflect the property of the system defined by
(\ref{action}) and (\ref{int}) that the equations of motion for
spinors are of first order in the evolution parameter. They are
obviously of second class:
\begin{equation}
\label{secondclasscon1}
     \{G_{I}^{\alpha},G_{J}^{\beta}\} = 2\delta_{IJ}C^{\alpha \beta}\;.
\end{equation}
There is in addition the kinematic constraint related to the
reparametrization invariance of the corresponding action functional.
It is given by canonical Dirac hamiltonian:
\begin{equation}
\label{diracham}
    H_{\rm D} = \frac{1}{2}(p^2 + m_0^2) -
    \sum_{I=1}^{2}\frac{h}{2}(-1)^{I}p_{\mu}J^{\mu}_{I}\;,
\end{equation}
where $p_{\mu}$ are the momenta canonically conjugated to the
positions $x^{\mu}$. The currents $J_{I}^{\mu}$ are bilinear in
spinors and their conjugated momenta: $J_{I}^{\mu} =
\pi_{I}^{\alpha}(\gamma^{\mu})_{\alpha\beta}\eta_I^{\beta}$. The
1-bein variable
was eliminated by putting $e=1$. \\
The constraints (\ref{secondclasscon}) together with
(\ref{diracham}) constitute the closed system of mixed type as one
has:
\begin{equation}
\label{poissonmixed}
    \{H_{\rm D},G_{I}^{\alpha} \} = (-1)^{I}\frac{h}{2}p_{\;\beta}^{\alpha}G_{I}^{\beta}\;.
\end{equation}
It should be stressed that the expression (\ref{diracham}) and the
relation (\ref{poissonmixed}) are so simple exactly due to the
substraction of the fourth order term (\ref{int}) $\sim j_{1}\cdot
j_{2}$ from
the Lagrange function (\ref{action}). \\
\\
As it was noted in \cite{DHW} the system of constraints
(\ref{secondclasscon}) and (\ref{poissonmixed}) can be replaced by
an equivalent system of first class by polarization of
(\ref{secondclasscon1})\cite{GB}. For massive momenta $(p^2<0)$ the
polarization of constraints is necessarily complex. \\
For this reason the system is most conveniently described in terms
of minimal building blocks of $\mathbf{Spin}(3,1) \approx \mathbf{SL}(2;\mathbb{C})$
representations i.e. in terms of complex Weyl spinors. \\
The real space of Majorana spinors $(\eta_I^{\alpha},
\pi_I^{\alpha})$ decomposes into, mutually complex
conjugated\footnote{According to common convention $z_I^{\bar{A}} =
\overline{({z}_I^A)}$.}, Weyl components $(z_I^A,\mathfrak{z}_I^A
)_{A=1,2}$ and $(z_{I}^{\bar{A}},\mathfrak{z}_{I}^{\bar{A}})
_{\bar{A} = 1,2}\;$. They span the eigensubspaces of $\gamma^5$
matrix corresponding to $\pm
i$ eigenvalues. \\
The Poisson brackets of the canonical Weyl variables are given as
follows:
\begin{equation}
\label{poissoncomplex}
    \{\mathfrak{z}_I^A,z_J^B\} =
\epsilon^{AB}\delta_{IJ}\;\;,\;\;\{\mathfrak{z}_I^{\bar{A}},z_J^{\bar{B}}\}
= \epsilon^{{\bar{A}}{\bar{B}}}\delta_{IJ}\;,
\end{equation}
where $\epsilon^{AB}$ $(\epsilon^{{\bar{A}}{\bar{B}}})$ are
$\mathbf{SL}(2;\mathbb{C})$ invariant tensors. \\
The second class constraints of (\ref{secondclasscon}) relate the
canonical Weyl coordinates:
    $G_I^A = \mathfrak{z}_I^{{A}} + z_I^A = 0$ and $ G_I^{\bar{A}} =
\mathfrak{z}_I^{\bar{A}} + z_I^{\bar{A}}=0$. Their polarized
counterparts can be expressed in the following way:
\begin{equation}
\label{GWp}
    G^{A}_{I(\pm)} = p^A_{\;\bar{B}}G_{I}^{\bar{B}} \pm im(p)
G_{I}^{A}\;\;,\;\;G^{\bar{A}}_{I(\pm)} =
p^{\bar{A}}_{\;{B}}G_{I}^{{B}} \pm im(p) G_{I}^{\bar{A}}\;,
\end{equation}
where $p^A_{\;\bar{B}}$ and $p^{\bar{A}}_{\;{B}}$ are (mutually
complex adjoint) matrix elements of the real operator
$p^{\mu}\gamma_{\mu}$ in the complex basis of Weyl spinors. Due to
Clifford algebra relations they do satisfy:
$p^A_{\;\bar{B}}p^{\bar{B}}_{\; {C}} = p^2\delta^A_C\,$ and
$p^{\bar{A}}_{\;{B}}p^{{B}}_{\; \bar{C}} =
p^2\delta^{\bar{A}}_{\bar{C}}\,$.
\\
The Hamiltonian constraint rewritten in terms of Weyl variables
takes the form\footnote{All indices are raised and lowered by
$\mathbf{SL}(2;\mathbb{C})$ invariant tensors $\epsilon^{AB}$
$(\epsilon^{{\bar{A}}{\bar{B}}})$ and their inverses $\epsilon_{AB}$
$(\epsilon_{{\bar{A}}{\bar{B}}})$ respectively.}:
\begin{equation}
\label{kinetic}
    H_{D} = \frac{1}{2}(p^2 + m_0^2) -
\sum_{I=1}^{2}\frac{h}{2}(-1)^I(\mathfrak{z}_I^Ap_{A\bar{B}}z_I^{\bar{B}}
+ \mathfrak{z}_I^{\bar{A}}p_{\bar{A}B}z_I^B)\;.
\end{equation}
The Poisson algebra of the complex constraints can be easily
calculated. From (\ref{poissoncomplex}) and (\ref{GWp}) it follows
that:
\begin{equation}
\label{zero}
    \{G^{A}_{I(\pm)},G^{B}_{J(\pm)}\} = 0 \;,\;\;\{G^{A}_{I(+)},G^{{B}}_{J(-)}\}=4\delta
_{IJ}m^2(p)\epsilon^{A {B}}\;.
\end{equation}
The functions (\ref{GWp}) are, under the Poisson bracket, the
mass-weighted eigenfunctions of (\ref{kinetic}):
\begin{equation}
\label{masseigen} \{H_D,G^{A}_{I(\pm)}\} = \pm \frac{ih}{2}(-1)^I
m(p)\,G^{A}_{I(\pm)}\;.
\end{equation}
As one should expect, the constraints $G^{A}_{I(\pm)}$ and
$G^{\bar{A}}_{I(\pm)}$ are not independent:
\begin{equation}
\label{gg}
    {G_{I(\pm)}^{\bar{A}}} =
    \mp\frac{i}{m(p)}p^{\bar{A}}_{\,{B}}G^{{B}}_{I(\pm)}\;.
\end{equation}
Hence, according to (\ref{zero}) and (\ref{masseigen}), the systems
$(H_D,G^{A}_{I(\pm)})$ constitute polarized Poisson algebras of
first class.  From $\overline{({G^{{A}}_{I(\pm)}})} =
G_{I(\mp)}^{\bar{A}}$ and (\ref{gg}) it follows that they are
mutually complex conjugated.

\section{The quantum model}

\subsection{The space of states}
The Hilbert space of states corresponding to the classical system
under consideration, consists in square integrable functions of
space-time momentum $(p^\mu)$ and Weyl spinor coordinates
$(z_{I}^A,z_{I}^{\bar{A}})$. In this representation space the
canonically conjugated variables are realized as the following
operators: $\;{\hat x}^{\mu} = -i{\partial}/{\partial p_{\mu}}$ and
$\; \hat {\mathfrak{z}}_{I}^A = i\epsilon^{AB}{\partial}/{\partial
z_{I}^B}$,$\;\; \hat {\mathfrak{z}}_{I}^{\bar{A}} =
i\epsilon^{\bar{A}\bar{B}}{\partial}/{\partial z_{I}^{\bar{B}}}\;$.
\\
It should be stressed that according to the conditions imposed
already at the classical level, the momenta $(p^\mu)$ are restricted
to the open domain $p^2<0$ supporting real massive particles. Since
this domain is the union of two disjoint cone interiors $p^0>0$ and
$p^0<0$, the Hilbert space of states decomposes into the
corresponding direct sum:
\begin{equation}
\label{sum}
    H = H^{\uparrow} \oplus H^{\downarrow}\;.
\end{equation}
The above direct components contain the functions of the
future-pointed
and past-pointed momenta respectively. \\
In the representation space above the operators corresponding to the
spinorial constraints (\ref{GWp}) are expressed by:
\begin{equation}
\label{qconstraints}
    G^{A}_{I(\pm)}= ip^{A {\bar{B}}}\frac{\partial}{\partial
    z_{I}^{\bar{B}}}
    \mp m(p)\epsilon^{AB}\frac{\partial}{\partial z_{I}^B} +
    p^A_{\;\;\bar{B}}z_{I}^{\bar{B}} \pm im(p) z_{I}^{A}\;,
\end{equation}
while the Dirac Hamiltonian (\ref{kinetic}) takes the following
operator form:
\begin{equation}
\label{qham}
     H_{D} = \frac{1}{2} \left (p^2 + m_0^2 \right) -
     \frac{ih}{2}\sum_{I=1}^{2} (-1)^I \left ({z}_{I}^{\bar{B}}p_{\bar{B}}^{\;\;\;A}
     \frac{\partial}{\partial
    z_{I}^{A}}
    + {z}_{I}^{{B}}p_{B}^{\;\;\;\bar{A}}\frac{\partial}{\partial
    z_{I}^{\bar{A}}} \right )\;.
\end{equation}
In order to analyze the spectrum of the model it is useful to
distinguish the differential part of (\ref{qham}):
\begin{equation}
\label{diff} S = -\frac{ih}{2}\sum_{I=1}^{2} (-1)^I \left
({z}_{I}^{\bar{B}}p_{\bar{B}}^{\;\;\;A}\frac{\partial}{\partial
z_{I}^{A}}
    + {z}_{I}^{{B}}p_{B}^{\;\;\;\bar{A}}\frac{\partial}{\partial z_{I}^{\bar{A}}} \right)\;.
\end{equation}
As it will appear clear the operator (\ref{diff}) is responsible for
spin-mass correlation in the spectrum of the model. For this reason
it will be called
the spin-mass coupling operator. \\
\\
The space of physical states of the model is defined as a kernel of
either $G^{A}_{I(+)}$ or $G^{A}_{I(-)}$ constraints\cite{GB}
(off-shell
physical states) and $H_{D}$ kinematical constraint. \\
According to \cite{Lop} and the detailed analysis of the paper
\cite{DHW} the physical states should be looked for within the
vectors of the form:
\begin{equation}
\label{states}
    \Psi_{(\pm)}(p,z_{I},\bar{z}_{I}) = W(z_{I},\bar{z}_{I})\,\Omega_{(\pm)} (p) \;,
\end{equation}
where $W(z_{I},\bar{z}_{I})$ are the polynomials in the Weyl
variables with momentum dependent coefficients.
\\
The space (\ref{sum}) of the functions (\ref{states}) splits in a
natural way into the sum of the direct integrals:
\begin{equation}
\label{directint} {H}^{\uparrow / \downarrow} =
\int\limits_{V_{(\pm)}} d^4p~ {H}^{\uparrow / \downarrow}(p)\;,
\end{equation}
of the Hilbert spaces ${H}^{\uparrow}(p)$  and ${H}^{\downarrow}(p)$
localized at the momentum $p$, contained in $V_{(\pm)}$ - the
future-pointed and the past-pointed cone
interiors respectively. \\
The states $\Omega_{(\pm)} (p)$ (the spin vacua) are the Gaussian
solutions of spinorial constraints $G^{A}_{I(\pm)}\Omega_{(\pm)} (p)
=0$. It is easy to check that (up to multiplicative factors) they
are given by:
\begin{equation}
\label{vacuum}
    \Omega_{(\pm)} (p) = \exp \left (\pm \sum_{I=1}^{2}
    \frac{z_{I}^{\bar{A}}p_{\bar{A}B}z_{I}^B}{m(p)} \right )\;.
\end{equation}
According to the adopted conventions the momentum matrix
$(p_{\bar{A}B})$ is negatively defined in the future-pointed
$(p^0>0)$ light-cone interior and it is positive in past-pointed
$(p^0<0)$ domain. For this reason one should take $\Omega_{(+)}(p)$
spin vacuum in $H^{\uparrow}(p)$ and $\Omega_{(-)}(p)$ state in
$H^{\downarrow}(p)$. This choice guarantees that the states
(\ref{states}) belong to localized Hilbert space of the quantum
model, i.e. they are square integrable with respect to Weyl
variables. It also means, that any state from $H^{\uparrow /
\downarrow}(p)$ can be represented as a superposition of the
following vectors:
\begin{eqnarray}
\label{degspinstate}
    \Psi_{(j_1,j_2)}^{\uparrow / \downarrow}(p,z_{I},\bar{z}_{I})
    &=&
    \sum\limits_{n_{1}=0}^{2j_1}\sum\limits_{n_{2}=0}^{2j_2} \Psi_{A_{1}^1\ldots
A_{2j_1-n_{1}}^1\bar{B}_{1}^1\ldots\bar{B}_{n_{1}}^1;\,A_{1}^2\ldots
A_{2j_2-n_{2}}^2\bar{B}_{1}^2\ldots\bar{B}_{n_{2}}^2} (p) ~\cdot \cr
&~& \cr &\cdot& z_{1}^{A_{1}^1}\cdots z_{1}^{{A}^{1}_{2j_1-n_{1}}}
z_{1}^{\bar{B}_{1}^1}\cdots z_{1}^{\bar{B}_{n_{1}}^1}
z_{2}^{A_{1}^2}\cdots z_{2}^{A_{2j_2-n_{2}}^2}
z_{2}^{\bar{B}_{1}^2}\cdots z_{2}^{\bar{B}_{n_{2}}^2 } ~\cdot \cr
&\cdot&\Omega_{(\pm)} (p) \;,
\end{eqnarray}
of fixed integral degrees $(2j_1,2j_2)$ in $(z_1^{A},z_1^{\bar{A}})$
and
$(z_2^{A},z_2^{\bar{A}})$ complex variables respectively. \\
In the light of the above considerations one should impose
$G^{A}_{I(+)}$ constraints in $H^{\uparrow}$ and $G^{A}_{I(-)}$ in
$H^{\downarrow}$. Hence, the space $\widehat{H}$ of physical
off-shell states is necessarily of the form of the following direct
sum:
\begin{equation}
\label{sum1}
  \widehat{H} = \widehat{H}_{(+)}^{\uparrow} \oplus \widehat{H}_{(-)}^{\downarrow}\;,
\end{equation}
where $\widehat{H}^{\uparrow / \downarrow}_{(\pm)}$ are the kernels
of $G^{A}_{I(\pm)}$ operators correspondingly. \\
In order to find these kernels one may follow the direct method
presented in the paper \cite{DHW}. They are however much more
transparently described in terms of spectrum generating algebra.
This algebra contains the constraints operators (\ref{qconstraints})
and in addition the following ones:
\begin{equation}
\label{dconstraints}
    D^{A}_{I(\pm)}= ip^{A {\bar{B}}}\frac{\partial}{\partial
    z_{I}^{\bar{B}}}
    \pm m(p)\epsilon^{AB}\frac{\partial}{\partial z_{I}^B} -
    p^A_{\;\;\bar{B}}z_{I}^{\bar{B}} \pm im(p) z_{I}^{A}\;.
\end{equation}
The above operators commute with these of (\ref{qconstraints}). The
only non-zero commutators are given by:
\begin{equation}
\label{com} [\;G^{A}_{I(+)},G^{\bar{B}}_{J(-)}\;]=4\delta
_{IJ}m(p)p^{A \bar{B}}\;\;,\;\;
[\;D^{A}_{I(+)},D^{\bar{B}}_{J(-)}\;]=4\delta _{IJ}m(p)p^{A
\bar{B}}.
\end{equation}
Since one has $(G^{A}_{I(\pm)})^{*}=G^{\bar{A}}_{I(\mp)}$ and
$(D^{A}_{I(\pm)})^{*}=D^{\bar{A}}_{I(\mp)}$, the operators
(\ref{qconstraints}) and (\ref{dconstraints}) may serve as creation
and anihilation operators
in the space (\ref{directint}). \\
From the form of the generators of $\mathbf{SL}(2;\mathbb{C})$
transformations\footnote{They can be obtained as operator
counterparts of the corresponding Noether conserved quantities
calculated from (\ref{action}) and (\ref{int}).}:
\begin{equation}
\label{lorentz}
    L^{\mu\nu} = i\left(p^\mu \frac{\partial}{\partial p^\nu} - p^\nu \frac{\partial}{\partial p^\mu}
    \right)+ \frac{i}{2}\sum_{I=1}^{2}\left( z_{I}^A \sigma ^{(\mu\nu) B}_{\;A} \frac{\partial}{\partial z_{I}^B} +
    z_{I}^{\bar{A}} \sigma ^{(\mu\nu) {\bar{B}}}_{\;{\bar{A}}} \frac{\partial}{\partial
    z_{I}^{\bar{B}}}\right)\;,
\end{equation}
it follows that the constraints $G^{A}_{I(\pm)}$ and the operators
$D^{A}_{I(\pm)}$ are of spinorial character and they carry spin
$\frac{1}{2}$:
\begin{equation}
\label{spinorial1}
[\;L^{\mu\nu},G^{A}_{I(\pm)}\;]=\frac{i}{2}G^{B}_{I(\pm)}\sigma
^{(\mu\nu) {{A}}}_{\;{{B}}} \;\;,\;\;
[\;L^{\mu\nu},D^{A}_{I(\pm)}\;]=\frac{i}{2}D^{B}_{I(\pm)}\sigma
^{(\mu\nu) {{A}}}_{\;{{B}}}\;,
\end{equation}
while the kinematic constraint:
\begin{equation}
\label{hamiltonian}
    H_{D} =\frac{1}{2} \left (p^2 + m_0^2 \right) -
     \frac{h}{8m^2(p)}\sum_{I=1}^{2} (-1)^I \left (G^{A}_{I(+)}p_{A\bar{B}}
     G^{\bar{B}}_{I(-)} -D^{A}_{I(+)}p_{A\bar{B}}
     D^{\bar{B}}_{I(-)}
    \right )\;,
\end{equation}
 is scalar. \\
One may directly check that:
\begin{equation}
\label{vac} G^{A}_{I(\pm)}\Omega_{(\pm)} (p) =
D^{A}_{I(\pm)}\Omega_{(\pm)}(p) = 0\;,
\end{equation}
and consequently, any state (\ref{degspinstate}) from the space
${H}^{\uparrow / \downarrow}(p)$ can be equivalently and more
transparently expressed in the Fock-type representation:
\begin{equation*}
    \Psi_{(j_1,j_2)}^{\uparrow / \downarrow}(p)
    =
    \sum\limits_{n_{1}=0}^{2j_1}\sum\limits_{n_{2}=0}^{2j_2}
    \Phi(p)_{A_{1}..
A_{2j_1-n_1}B_1 .. B_{n_1}C_1 .. C_{2j_2-n_2}D_{1}.. D_{n_2} (p)}
~\cdot \;\;\;\;\;\;\;\;\;\;\;\;\;\;\;\;\nonumber
\end{equation*}
\begin{equation}
\label{opstate} \;\;\;\;\;\;\;\;\;\cdot D_{1(\mp)}^{A_{1}}..
D_{1(\mp)}^{{A}_{2j_1-n_{1}}} G_{1(\mp)}^{B_{1}}..
G_{1(\mp)}^{{B}_{n_{1}}} D_{2(\mp)}^{C_{1}}..
D_{2(\mp)}^{C_{2j_2-n_{2}}}G_{2(\mp)}^{{D}_{1}}..
G_{2(\mp)}^{{D}_{n_{2}} }\Omega_{(\pm)}(p) \;.
\end{equation}
As one has $[G^{A}_{I(+)},G^{{B}}_{J(-)}]=4i\delta
_{IJ}m^2(p)\epsilon^{A {B}}$ and
$[D^{A}_{I(+)},D^{{B}}_{J(-)}]=4i\delta _{IJ}m^2(p)\epsilon^{A
{B}}$, the constraints are easily solvable now. They simply mean
that $G_{I(\mp)}^{A}$ excitations should be absent in
(\ref{opstate}) and for the solutions one gets:
\begin{eqnarray}
\label{opstate1}
    \Psi_{(j_1,j_2)}^{\uparrow / \downarrow}(p)
    =
    \Phi(p)_{{A_{1}}..
A_{2j_1}C_{1}.. C_{2j_2}}D_{1(\mp)}^{A_{1}}..
 D_{1(\mp)}^{{A}_{2j_1}}
D_{2(\mp)}^{C_{1}}..D_{2(\mp)}^{C_{2j_2} } \Omega_{(\pm)} (p) \;.
\end{eqnarray}
It is worth to mention that the constraints $G_{I(\pm)}^{A}$ imposed
on the states in the form of (\ref{degspinstate}) become the
Dirac-type equations for the coefficients of the polynomials. In the
case of ${\widehat H}^{\uparrow}_{(+)}(p)$ they take the following
form:
\begin{eqnarray}
\label{equation} &-& (n_I+1)
p_{A_{2j_{I}-n_{I}}^{I}}^{\bar{B}_{n_{I}+1}^I}\Psi_{A_{1}^{1}\ldots
    A_{2j_{1}-n_{1}-
    \delta_{I,1}}^{1}\bar{B}_{1}^{1}\ldots\bar{B}_{n_{1}+\delta_{I,1}}^{1};\,A_{1}^{2}\ldots
    A_{2j_{2}-n_{2}- \delta_{I,2}}^{2}\bar{B}_{1}^{2}\ldots\bar{B}_{n_{2}+\delta_{I,2}}^{2}}(p) ~=\cr
    &~& \cr
    &=&
    i m(p)(2j_{I}-n_{I})\Psi_{A_{1}^{1}\ldots
    A_{2j_{1}-n_{1}}^{1}\bar{B}_{1}^{1}\ldots\bar{B}_{n_{1}}^{1};\,A_{1}^{2}\ldots
    A_{2j_{2}-n_{2}}^{2}\bar{B}_{1}^{2}\ldots\bar{B}_{n_{2}}^{2}}
    (p)\;.
\end{eqnarray}
where $I=1,2$ and $n_I = 0,...,2j_I-1$. \\
In the complementary space ${\widehat H}^{\downarrow}_{(-)}(p)$ of
anti-particles the above equations are completely analogous. \\
\\
    In order to find the spectrum of the model one should impose the
kinematical constraint $H_D$ (\ref{hamiltonian}) on the solutions of
(\ref{opstate1}). This constraint correlates the masses of the
particles with their spins. The detailed analysis of the spectrum
will be given in the next subsection.
\subsection{Spin and mass spectrum}

In order to find the spectrum explicitly it is convenient to use the
following decomposition of the off-shell spaces:
\begin{equation}
\label{decomposition} \widehat{H}_{(+)}^{\uparrow}(p) =
\bigoplus_{(j_1,j_2)} \widehat{H}_{(+)}^{\uparrow (j_1,j_2)}(p) \;,
\end{equation}
where $\widehat{H}_{(+)}^{\uparrow (j_1,j_2)}(p)$ contains the
polynomials of fixed bidegree $(2j_1,2j_2)$ of the form of
(\ref{opstate1}). From (\ref{hamiltonian}) and commutations
relations (\ref{com}) it follows that the spin-mass coupling
operator $S$ of (\ref{diff}) is diagonal and acts in the following
way:
\begin{equation}
\label{saction} S \widehat{H}_{(+)}^{\uparrow (j_1,j_2)}(p) =
{h}m(p) (j_1 - j_2) \widehat{H}_{(+)}^{\uparrow (j_1,j_2)}(p)\;.
\end{equation}
Hence, the kinetic constraint imposes the following mass-shell
condition on the particle momenta:
\begin{equation}
\label{momentum}
    \frac{1}{2}(p^2 + m_0^2) + {h}m(p)(j_1 - j_2) = 0\;,
\end{equation}
with the unique solution leading to normalizable states\footnote{The
second, negative root of (\ref{m+}) has to be rejected because it
makes the exponent (\ref{vacuum}) growing very fast at infinity.},
given by:
\begin{equation}
\label{m+} m_{(j_1,j_2)}^{\uparrow} = \sqrt{h^2(j_1-j_2)^2+m_0^2} +
h(j_1 - j_2) \;\;;\;\; j_1,~j_2\geq 0\;.
\end{equation}
One may analogously find the masses of the states supported by the
past-pointed cone interior:
\begin{equation}
\label{m-} m_{(j_1,j_2)}^{\downarrow} = \sqrt{h^2(j_1-j_2)^2+m_0^2}
- h(j_1 - j_2) \;\;;\;\; j_1,~j_2\geq 0\;.
\end{equation}
According to (\ref{m+}) and (\ref{m-}), for on-shell physical
states, the direct integrals corresponding to (\ref{directint}) are
restricted to the respective mass-shells. The spaces of physical
on-shell states will be denoted by $\widetilde{H}_{(+)}^{\uparrow}$
and
$\widetilde{H}_{(-)}^{\downarrow}$ respectively. \\
The comparison of the formulae (\ref{m+}) and (\ref{m-}) clearly
indicates that the spectrum of the model is CPT
invariant\cite{Jost}, i.e. it is symmetric with respect to
particle-antiparticle exchange. This property was missing in the
simple prototype model
presented in \cite{DHW}. \\
\\
In order to perform a more detailed analysis of the spectrum it is
useful to distinguish two families of subspaces of the on-shell
space $\widetilde{H}_{(+)}^{\uparrow}(p)$, namely:
\begin{equation}
\label{distin} \widetilde{H}_{(+)}^{\uparrow (j_1,0)}(p)\;\;\;\;
{\rm and}\;\;\;\; \widetilde{H}_{(+)}^{\uparrow (0,j_2)}(p) \;,
\end{equation}
where the numbers $j_1$ and $j_2$ have direct physical
interpretation as particle spins\cite{Lop}. As it is clearly visible
from (\ref{opstate1}) these spaces describe the states of single
particles with spins $j_1$ and $j_2$, and with the corresponding
masses:
\begin{equation}
\label{m+m-} m_{j_1}^{\uparrow} = \sqrt{h^2j_1^2+m_0^2} +
hj_1\;\;,\;\;m_{j_2}^{\uparrow} = \sqrt{h^2j_2^2+m_0^2} - hj_2\;.
\end{equation}
The antiparticles of $\widetilde{H}_{(+)}^{\uparrow (j_1,0)}(p)$ and
$\widetilde{H}_{(+)}^{\uparrow (0,j_2)}(p)$ are represented in the
spaces:
\begin{equation}
\label{distin1} \widetilde{H}_{(-)}^{\downarrow (0,j_1)}(p)\;\;\;\;
{\rm and}\;\;\;\; \widetilde{H}_{(-)}^{\downarrow (j_2,0)}(p)\;,
\end{equation}
respectively. \\
According to (\ref{directint}), as it was already mentioned, the
localized spaces (\ref{distin}) and (\ref{distin1}) fit together
into direct integrals over the corresponding mass-shells
${S_{+}(j_I)}$:
\begin{equation}
\label{h1h2} \widetilde{H}_{(+)}^{\uparrow (j_1,0)} =
\int\limits_{S_{+}(j_1)} d\mu_{j_1}(p) \widetilde{H}_{(+)}^{\uparrow
(j_1,0)}(p)\;\;\;\; {\rm and}\;\;\;\; \widetilde{H}_{(+)}^{\uparrow
(0,j_2)} = \int\limits_{S_{+}(j_2)} d\mu_{j_2}(p)
\widetilde{H}_{(+)}^{\uparrow (0,j_2)}(p)\;,
\end{equation}
with respect to Lorentz invariant measures $d\mu_{j_I}$.  \\
Together with their antiparticles partners they describe two
particle-antiparticle pairs located at two diverging Regge
trajectories. Hence, the corresponding mass-levels are
non-degenerate as in the case of the prototype model of \cite{DHW}.
For this reason these trajectories will be called elementary. They
are illustrated at Fig.1A\footnote{For $p^0<0$ one obtains exactly
the same trajectories of antiparticles.}. \\
\\
Besides of the particle pairs located at the elementary
trajectories, there is a family of Hilbert spaces containing the
polynomials non-zero degree in both, $(z_1^{A},z_1^{\bar{A}})$ and
$(z_2^{A},z_2^{\bar{A}})$ Weyl variables, i.e. of the spaces:
\begin{equation}
\label{composit} \widetilde{H}_{(+)}^{\uparrow (j_1,j_2)}(p)\;\;\;\;
{\rm where}\;\;\;\;j_{I} > 0\;\;;\; I=1,2\;.
\end{equation}
The momenta of the states from (\ref{composit}) are obviously
constrained by mass-shell condition (\ref{m+}). The corresponding
antiparticles are contained in $\widetilde{H}_{(-)}^{\downarrow
(j_2,j_1)}(p)$ and according to (\ref{m-})
they have exactly this same masses. \\
In contrast to $\widetilde{H}_{(+)}^{\uparrow (j_1,0)}$ and
$\widetilde{H}_{(+)}^{\uparrow (0,j_2)}$ the spaces (\ref{composit})
are reducible with respect to $\mathbf{SL}(2;\mathbb{C})$ group
representation, i.e. they decay into direct sums of the subspaces
containing the states with fixed spins, i.e. those of elementary
particles. This property is clearly visible from the form of the
solution of $G^{A}_{I(+)}$ constraints (\ref{qconstraints}) which is
explicitly given in (\ref{opstate1}). For this reason it is natural
to call the states from (\ref{composit})
the clusters. \\
The spin content of the spaces $\widetilde{H}_{(+)}^{\uparrow
(j_1,j_2)}(p)$ at fixed momentum can be obtained by the standard
simple rule of the decomposition of $\mathbf{SU}(2)$ (the little
group of massive momentum $p$) representations\cite{JM}:
\begin{equation}
\label{cg} \widetilde{H}_{(+)}^{\uparrow (j_1,j_2)}(p) =
\bigoplus_{j=|j_1-j_2|}^{j_1+j_2}\widetilde{H}_{(+)}^{\uparrow
j}(p)\;,
\end{equation}
where $\widetilde{H}_{(+)}^{\uparrow j}(p)$ denote the irreducible
subspace of spin $j$. \\
From (\ref{cg}) it is clear that keeping the mass of the particle
fixed (fixed value of the difference $j_1-j_2$) one may obtain
arbitrarily high spin states. Hence, the states from
$\widetilde{H}_{(+)}^{\uparrow (j_1,j_2)}(p)$ generate infinite spin
degeneracy of the mass levels. The corresponding cluster
trajectories are illustrated on Fig.1B. From this diagram it is
clear that the cluster states contribute to elementary trajectories
and in addition, they generate their own ones.
\begin{figure}[h]
\begin{center}
\includegraphics[width=\textwidth, height=12cm]{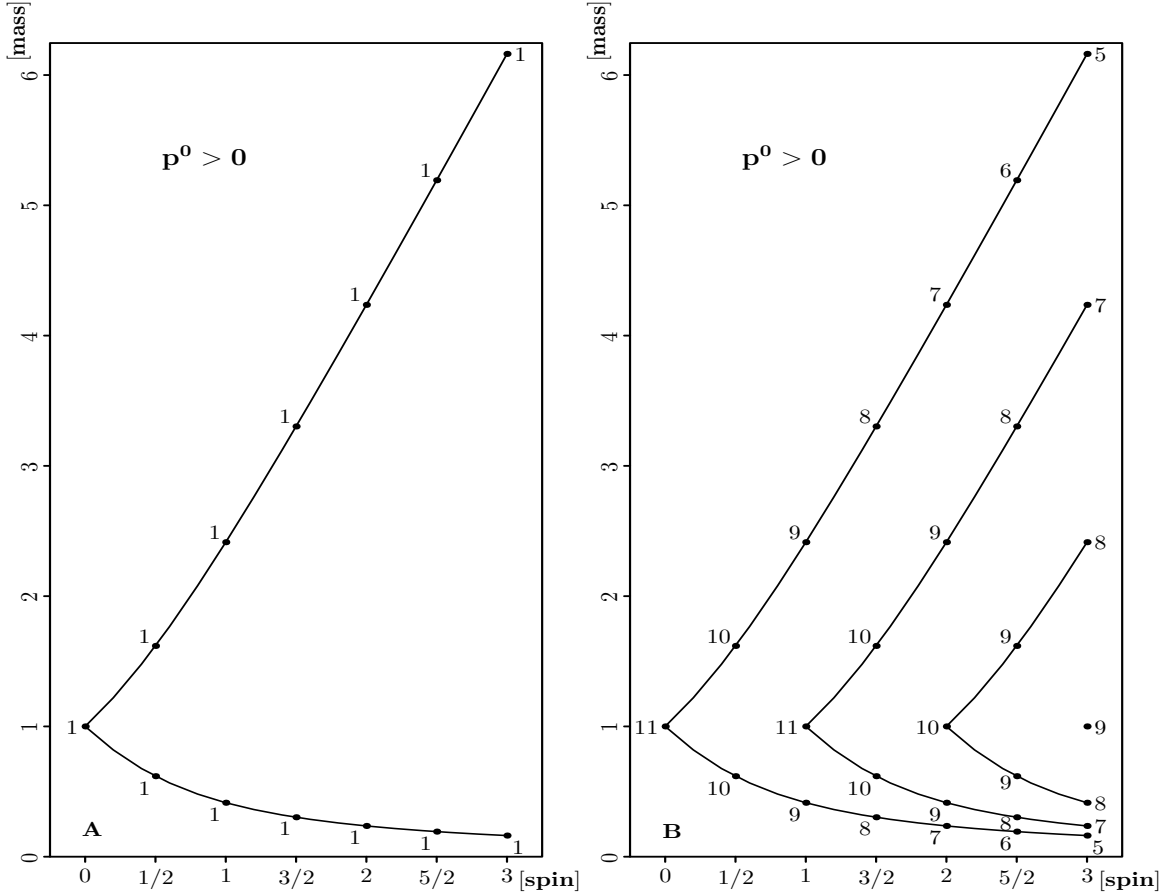}
\end{center}
\caption{{\bf A} - The elementary trajectories for particles (in
$m_0$ units). {\bf B} - The degeneracy of the spectrum of clusters
(in $m_0$ units and with $j_I$ ranging from $0$ to $11/2$).}
\label{fig.1}
\end{figure}\\

\section*{Conclusions and outlook}

As it was shown the model considered in this paper has CPT invariant
mass spectrum. It consists of elementary (non-degenerate) Regge
trajectories and highly degenerate sea of cluster particles.
\\
According to the considerations of second chapter, in the case of
elementary trajectories, the solution (\ref{opstate1}) describes the
wave function of the one elementary particle with fixed spin (in the
sense of $\mathbf{SL}(2;\mathbb{C})$ representations). The particle
is
accompanied by the corresponding antiparticle. \\
In the case of cluster (\ref{cg}) the wave functions solving the
constraints equations (\ref{qconstraints}), (\ref{qham}) do not
represent the elementary objects, but the families of particles with
spins ranging from $|j_1-j_2|$ to $j_1+j_2$. The solution
(\ref{opstate1}) together with mass-shell condition
describe an ensemble of particles. \\
These objects make an impression as being untypical, however the
idea of cluster fields (in that case with respect to mass not spin)
was proposed and developed almost 50 years ago in the series of
papers \cite{clusters}. \\
\\
The  analysis at the classical level indicates that the infinite
degeneracy of the mass levels might appear as an effect of the
substraction of the interaction term (\ref{int}) from
(\ref{action}), which makes the classical dynamics of the
corresponding spinors linear, and almost completely independent.
This substraction results in the independence of the polarized
constraints (\ref{GWp}) for both spinors and simple form of Dirac
Hamiltonian. For this very reason, the corresponding wave functions
contain products of independent polynomials in both independent Weyl
spinors, which in turn amounts to the decomposition
given in (\ref{cg}). \\
The detailed analysis of the model described by (\ref{action}), i.e.
containing the non-linear interaction (\ref{int}) is
already in progress, and will be presented in the future paper.\\

\section*{ Acknowledgements}
One of the authors (M.D.) would like to thank the friends and
colleagues from the Institute of Theoretical Physics in Bia{\l}ystok
for warm hospitality. \\
The authors would like to thank also prof. J.Lukierski, prof. Z.Haba and  dr. A.B{\l}aut
 for interesting discussions. \\
Special thanks (from Z.H.) are due to Bo\.{z}ena Gonciarz for her
inspiration. \\
This work is partially supported by KBN grant 1P03B01828.


\begin{thebibliography}{99}
\bibitem{BZ} A.O.Barut, N.Zoughi, Phys. Rev. Lett. 52 (1984) 2009
\bibitem{GT} V.D.Gershun, V.I.Tkach, JETP Lett. (1979) 320
\bibitem{HPPT} P.S.Howe, S.Penati, M.Pernici, P.Townsend, Phys. Lett.
B215 (1988) 255
\bibitem{LF} S.Fedoruk, J.Lukierski, Phys. Lett. B632 (2006)
371-378
%
\bibitem{LF1} A.Bette, J.de Azcarraga, J.Lukierski, C.Miquel-Espanya,
Phys. Lett. B595 (2004)
\bibitem{andrzej}A.Frydryszak, {\it Lagrangian models of particles
with spin}, Published in {\it From field theory to quantum groups},
Singapore World Scientific Publishing (1996)
\bibitem{KLS} S.M.Kuzenko, S.L.Lyakhovich, A.Yu.Segal,
Int. J. Mod. Phys. A10 (1995) 1529
\bibitem{HDS} Z.Hasiewicz, F.Defever, P.Siemion,
Int. J. Mod. Phys. A 7 (1992) 3979-3996
\bibitem{DHW} M.Daszkiewicz, Z.Hasiewicz, C.Walczyk, {\it High spin particles with spin-mass
coupling}, Advances in Applied Clifford Algebras (in press)
\bibitem{Jost}R.F.Streater, A.S.Wightman {\it PCT Spin - Statistic
and all that} Benjamin, New York 1964 \\
R.Jost {\it The General Theory of Quantized Fields} AMS, Providence,
Rhode Island 1965
\bibitem{schroedinger} E.Schroedinger, Sitzugsh. Preuss. Akad. Wiss.
Phys.-Math. Kl. 24 (1930) 418; 3 (1931) 1
\bibitem{BB} A.O.Barut, A.J.Bracken, Phys. Rev. D23 (1981) 2454
\bibitem{rivas} M.Rivas, {\it Classical elementary particles, spin, zitterbewegung and all that} -
physics/0312107
\bibitem{GB} S.Gupta, Proc. Roy. Soc. A63 (1950) 681 \\
K.Bleuler, Helv. Phys. Acta 23 (1950) 567
\bibitem{Lop}
Jan {\L}opusza\'{n}ski {\it Spinor Calculus} PWN Warsaw 1985 (in Polish) \\
A.O.Barut, R.R\c{a}czka {\it Theory of Group Representations and
Applications} PWN Warsaw 1977
\bibitem{JM} Jan Mozrzymas {\it Application of Group Theory in
Physics} PWN Warsaw 1976 (in Polish) \\
M.Hamermesh {\it Group Theory} PWN Warsaw 1968
\bibitem{field1} A.R.Swift, Phys. Rev. 176 (1968) 1848-1855
\bibitem{field2} W.J.Zakrzewski, Nuovo Cim. A60 (1969) 263-290
\bibitem{field3} A.R.Swift, R.W.Tucker, Nuovo Cim. A67 (1970) 345-355
\bibitem{field4} A.R.Swift, R.W.Tucker, Phys. Rev. D1 (1970) 2894-2900
\bibitem{Rebbi} C.Rebbi,  Phys. Rept. 12 (1974) 1-73
\bibitem{hikko} H.Hata, K.Itoh, T.Kugo, H.Kunitomo, K.Ogawa, Phys. Rev.
D34 (1986) 2360
\bibitem{witten} E.Witten, Nucl. Phys. B268 (1986) 253
\bibitem{Polyakov} A.M.Polyakov, Phys. Lett. B103 (1981) 207
\bibitem{clusters} O.W.Greenberg, Ann. of Phys. 16 (1961) 158 \\
A.L.Licht, Ann. of Phys. 34 (1965) 161 \\
L.Turko, Nucl. Phys. B114 (1976) 535-545

\end{thebibliography}
\end{document}